\documentclass[aps,prc,showpacs,twocolumn,superscriptaddress]{revtex4}
\usepackage{epsfig}
\begin{document}


\title{Experimental signature of collective enhancement in nuclear level density}

\author{Deepak Pandit}
\email[e-mail:]{deepak.pandit@vecc.gov.in}
\affiliation{Variable Energy Cyclotron Centre, 1/AF-Bidhannagar, Kolkata - 700064, India}

\author{Srijit Bhattacharya}
\affiliation{Department of Physics, Barasat Govt. College, Barasat, N 24 Pgs, Kolkata - 700124, India}

\author{Debasish Mondal}
\affiliation{Variable Energy Cyclotron Centre, 1/AF-Bidhannagar, Kolkata - 700064, India}
\affiliation{Homi Bhabha National Institute, Training School Complex, Anushaktinagar, Mumbai - 400094, India}

\author{Pratap Roy}
\affiliation{Variable Energy Cyclotron Centre, 1/AF-Bidhannagar, Kolkata - 700064, India}
\affiliation{Homi Bhabha National Institute, Training School Complex, Anushaktinagar, Mumbai -  400094, India}

\author{K Banerjee}
\email[Present address: Department of Nuclear Physics, Research School of Physics and Engineering, Canberra, ACT 2601, Australia]{}
\affiliation{Variable Energy Cyclotron Centre, 1/AF-Bidhannagar, Kolkata - 700064, India}
\affiliation{Homi Bhabha National Institute, Training School Complex, Anushaktinagar, Mumbai -  400094, India} 

\author{S. Mukhopadhyay}
\affiliation{Variable Energy Cyclotron Centre, 1/AF-Bidhannagar, Kolkata - 700064, India}
\affiliation{Homi Bhabha National Institute, Training School Complex, Anushaktinagar, Mumbai -  400094, India}

\author{Surajit Pal}
\affiliation{Variable Energy Cyclotron Centre, 1/AF-Bidhannagar, Kolkata - 700064, India}

\author{A. De}
\affiliation{Department of Physics, Raniganj Girls' College, Raniganj - 713358, India}

\author{Balaram Dey}
\affiliation{Saha Institute of Nuclear Physics, 1/AF-Bidhannagar, Kolkata - 700064, India}

\author{S. R. Banerjee}
\affiliation{(Ex)Variable Energy Cyclotron Centre, 1/AF-Bidhannagar, Kolkata - 700064, India}


\date{\today}

\begin{abstract}
We present a probable experimental signature of collective enhancement in the nuclear level density (NLD) by  measuring the neutron and the giant dipole resonance (GDR) $\gamma$ rays emitted from the rare earth $^{169}$Tm compound nucleus populated at 26.1 MeV excitation energy. An enhanced yield is observed in both neutron and $\gamma$ ray spectra corresponding to the same excitation energy in the daughter nuclei. The enhancement could only be reproduced by including a collective enhancement factor in the Fermi gas model of NLD to explain the neutron and GDR spectra simultaneously. The experimental results show that the relative enhancement factor is of the order of 10 and the fadeout occurs at $\sim$ 14 MeV excitation energy, much before the commonly accepted transition from deformed to spherical shape. We also explain how the collective enhancement contribution changes the inverse level density parameter ($k$) from 8 to 9.5 MeV observed recently in several deformed nuclei.                

\end{abstract}
\maketitle

The atom, consisting of a tiny nucleus of protons and neutrons surrounded by a cloud of electrons, is responsible for nearly all the properties of matter that have shaped the world around us. Although, the atomic properties are governed by the electronic structure, its existence is decided by the nucleus. It is a complex quantal system which is held together by the strong nuclear force. The nucleus attains variety of configurations even if a small excitation energy is provided to it. The density of nuclear levels increases rapidly with increasing excitation energy \cite{beth36, bohr69}. Thus, statistical models are not only appropriate but essential for the comprehension and prediction of different nuclear decays at moderate and high excitation energies. One of the important ingredients of the statistical model is the nuclear level density (NLD) which is defined as the number of excited levels per unit excitation energy. 
The NLD has important contribution in the calculations of explosive nuclear burning in astrophysical environments such as nuclear reaction rates in nucleosynthesis and reliable estimates of nuclear abundance \cite{raus00, raus97} as well as 
in nuclear fission \cite{capo09}, multifragmentation \cite{bond85} and spallation reactions \cite{davi10}. 
It also provides important information about the nuclear thermodynamic properties such as temperature (T), entropy and heat capacity \cite{melb99}. 
The NLD is extracted experimentally from counting the levels, neutron resonance studies \cite{huiz72}, 
Oslo technique \cite{schi00}, two-step cascade method \cite{becv92}, beta-Oslo method \cite{spyr14}, 
$\gamma$-ray calorimetry \cite{ullm13} and particle evaporation spectra \cite{voin06}.
Theoretically, it has been characterized by phenomenological analytical expressions \cite{beth36, igna79, gilb65} as well as calculations based on different microscopic approaches \cite{gori08, grim13, ozen13, hung17}.

Apart from the intrinsic excitation, the nucleus also displays collective vibrational and rotational motion analogous to atomic and molecular physics. These collective degrees of freedom introduce new levels up to moderate excitation energies, and their contribution is described as collective enhancement factor in the NLD. The contribution of collectivity in the NLD $\rho(E^*,J)$ at excitation energy $E^*$ and angular momentum $J$ is expressed phenomenologically \cite{igna79} as,
\begin{equation}
\rho (E^*,J) = \rho_\textrm{\scriptsize{int}}(E^*, J) * K_\textrm{\scriptsize{coll}} \label{eqn1}
\end{equation}
where $\rho_\textrm{\scriptsize{int}}(E^*, J)$ is the intrinsic single particle level density and $K_\textrm{\scriptsize{coll}}$ is the collective enhancement factor. 
Although, the NLD is indispensible in the study of nuclear decay, the collective enhancement in the NLD is still not a well understood topic due to the lack of experimental data.
The magnitude and exact form of $K_\textrm{\scriptsize{coll}}$ still remains an open question. Several expressions for $K_\textrm{\scriptsize{coll}}$  exist in literature where the degree of enhancement varies from 10 to 100 {\cite{ozen13, koni08, hans83, jung98}.  On the other hand, the earlier experimental studies have produced contradictory results on the collective enhancement and its fadeout {\cite{ jung98, koma07}. 
Quite recently, our extensive studies on neutron evaporation from several deformed nuclei have established the fact that the fadeout of collectivity is related to the nuclear shape phase transition and occurs at an excitation energy in the region of 14 - 21 MeV \cite{prat13, bane17}.    
While, a sharp change in the value of inverse level density parameter ($k$), within the initial compound nuclear excitation energy interval of 32-37 MeV, is observed for all the deformed nuclei ($^{169}$Tm, $^{173}$Lu, $^{185}$Re), a weak effect is observed for the near spherical $^{201}$Tl nucleus \cite{bane17}. Therefore, if there is an enhancement and its fadeout is in the region 14 - 21 MeV, then that should be directly evident in both neutron and giant dipole resonance (GDR) $\gamma$ decay spectra from the highly deformed rare-earth nuclei.   

The GDR is another collective mode of excitation of the nuclei which can be understood macroscopically as the out-of-phase oscillation between the protons and neutrons \cite{hara01, gaar92}.
Microscopically, it is conceived as the coherent superposition of particle-hole excitations. 
It is an indispensable tool in nuclear structure physics and has been utilized recently to determine the  
ratio of shear viscosity ($\eta$) to entropy density (s) of finite nuclear matter \cite{deba17}.   
The GDR $\gamma$ emission occurs early in the decay of excited nuclei and also couples directly with the nuclear shape degrees of freedom. Thus, the investigation of its strength distribution should provide information about nuclear deformation and any enhanced yield will present an experimental signature of the collective enhancement in the NLD.

The experiment was performed using the alpha beams (E\textrm{$_{Lab}$} = 28 MeV) from the K-130 cyclotron at the Variable Energy Cyclotron Centre, Kolkata by bombarding $^{165}$Ho target. The compound nucleus $^{169}$Tm (ground state deformation $\beta$ $\sim$ 0.3 \cite{moll95}) was populated at 26.1 MeV excitation energy. The critical angular momentum for the reaction was 11$\hbar$. The high-energy GDR $\gamma$ rays were detected at 90$^\circ$ and 125$^\circ$ with respect to the
incident beam direction by employing the LAMBDA spectrometer \cite{supm07}, arranged in a 7x7 matrix, at a distance of 50 cm. The time-of-flight (TOF) technique was employed to discriminate the neutrons from the high-energy $\gamma$ rays. The pulse shape discrimination (PSD) technique was adopted to reject pile-up events in the individual detector elements by measuring the charge deposition over two time intervals (30 ns and 2 $\mu$s). However, the pile-up events are very few due to high granularity of the detector array LAMBDA \cite{supm07}. 
The 50-element low-energy $\gamma$ multiplicity filter \cite{dipu10} was used (in coincidence with the high-energy $\gamma$ rays) to estimate the angular momentum populated in the compound nucleus in an event-by-event mode as well as to get the fast start trigger for the TOF measurements. The filter was split into two blocks of 25 detectors each of which were placed on the top and bottom of a specially designed scattering chamber at a distance of 4.5 cm from the target in staggered castle type geometry. The neutron evaporation spectra were measured using two 5$^{\prime\prime}$ x 5$^{\prime\prime}$ liquid-scintillator (BC501A) \cite{bane09} detectors (in coincidence with the multiplicity filter) placed outside the scattering chamber at  120$^\circ$ and 150$^\circ$ with respect to the beam direction and at a distance of 150 cm from the target. The energy of the emitted neutrons was measured using the TOF technique whereas the neutron-$\gamma$  discrimination was achieved by both PSD and TOF. 
The time resolution of the neutron detectors was typically about 1.2 ns which give an energy resolution of about 0.9 MeV at 10 MeV for the present setup. To keep the background at a minimum level the beam dump was kept at 3 m away from the target and was well shielded with layers of lead and borated paraffin. 
The schematic view of the experimental setup is shown in Fig. \ref{setup}. 
The details of the GDR \cite{dipu10a, dipu12, bala14, deba16} and the neutron analyses \cite{prat13, bane17, bane12} have already been discussed in our earlier papers.

\begin{figure}
\begin{center}
\includegraphics[height=8.0 cm, width=8.5 cm]{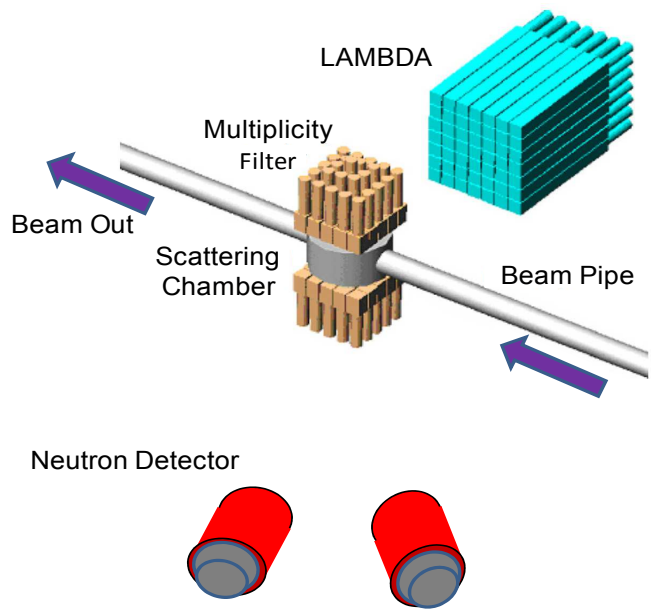}
\caption{\label{setup} (color online) Schematic view of the experimental setup.}
\end{center}
\end{figure}

\begin{figure}
\begin{center}
\includegraphics[height=11.0 cm, width=8.5 cm]{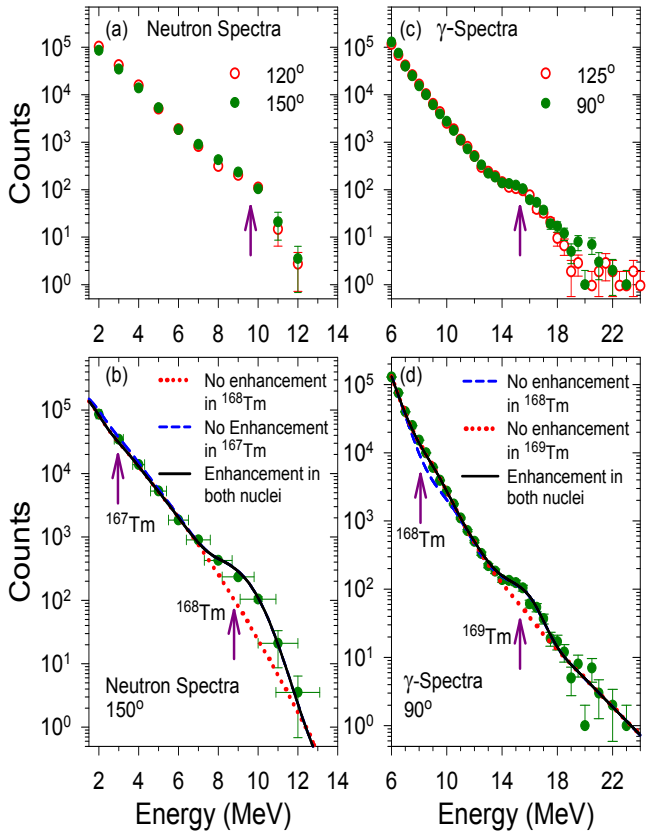}
\caption{\label{fig1} (color online) (a) The experimental neutron spectra measured at two angles are compared with each other. (b) The experimental neutron spectrum is compared with the CASCADE calculation. (c) The experimental $\gamma$-spectra measured at two angles are compared with each other. (d) The experimental $\gamma$-spectrum is compared with the CASCADE calculation plus bremsstrahlung component. The enhancement in the spectra and the contribution from different nuclei are shown with arrows.}
\end{center}
\end{figure}

The neutron and the high-energy $\gamma$ ray spectra, each measured at two different angles,
are shown in Fig. \ref{fig1} (a) and (c), respectively. As can be seen, the two spectra almost overlap with each other, which indicates that they have originated from an equilibrated compound nucleus. 
Most noteworthy is the large yield in both neutron energy spectrum (beyond 6 MeV) and GDR $\gamma$ ray (around 16 MeV) spectrum. This high energy GDR $\gamma$ ray at 16 MeV can only arise from fully energy equilibrated compound nucleus since the nonfusion events are accompanied by $\gamma$ rays less than 10 MeV \cite{drc12}. It is also interesting to note that the GDR and the neutron decay explore the same excitation energy region in the daughter nuclei $^{169}$Tm and $^{168}$Tm, respectively. In order to explain the experimental data, the neutron energy and the high-energy $\gamma$ ray spectra were calculated employing a modified version of statistical model code CASCADE \cite{srit08, srit14}. The shape of the particle spectra depends on the transmission coefficients of outgoing particles and the NLD of the residual nucleus. The transmission coefficients for statistical model calculation were obtained from the optical model where the potential parameters for neutron, proton and $\alpha$ were taken from Refs. \cite{pere76}, \cite{pere63} and \cite{mcfa66}, respectively. The experimental fold distribution measured using the 50-element $\gamma$-multiplicity filter was converted to the spin distribution through comparison with a GEANT simulation and was used as input for the calculation \cite{dipu10}. The intrinsic level density used in the modified version of CASCADE code is based on the Fermi gas model \cite{beth36} given as  
\begin{equation}
\rho_\textrm{\scriptsize{int}}(E^*, J) = \frac{2J+1}{12\theta^{3/2}} \sqrt{a} \frac{\exp{(2\sqrt{aU}})}{U^2}. \label{eqn2}
\end{equation}
Here $U$ = $E^*$ -  $\frac{J(J+1)}{2I_\textrm{\scriptsize{eff}}}$ - $\Delta_\textrm{\scriptsize p}$ is the available thermal energy. $\frac{J(J+1)}{2I_\textrm{\scriptsize{eff}}}$ is the energy bound in rotation and $\theta$ = $\frac{2I_\textrm{\scriptsize{eff}}}{\hbar^2}$, where $I_\textrm{\scriptsize{eff}}$ is the effective moment of inertia. The excitation energy is shifted back by the pairing energy $\Delta_\textrm{\scriptsize p}$ which is calculated using the relation $\Delta_\textrm{\scriptsize p}$ = $\frac{12}{\sqrt{A}}$. The NLD parameter $a$ is related to the single-particle density of states at the Fermi energy. The prescription of Ignatyuk \cite{igna75} was used for the level density parameter which is given as  
$a$=$\tilde{a}$[1+($\Delta$S/$U$)(1-exp(-$\gamma$$U$))] where $\tilde{a}$ = $A/k$, $\Delta$S is the shell correction and $\gamma$ is the shell damping factor. This parametrization takes into account the nuclear shell effects at low excitation energy and connects smoothly to the liquid drop value at high excitation energy and found
to explain the GDR data well \cite{srit14}. However, the shell correction factors for Tm isotopes are very small and less than 1.0 MeV \cite{moll95}.
 
It was observed that the variation in the transmission coefficients (using different prescriptions) and the deformation parameters were inconsequential. The shape of the neutron energy spectrum was determined by the inverse level density parameter only. Similarly, the $\gamma$ spectrum also depended only on the level density and the GDR parameters.
However, it was not possible to explain the enhanced yield obtained in both neutron and $\gamma$ spectra by changing the $k$ value and the GDR parameters even after taking into account the shell and pairing effects in level density. Therefore, in order to explain the experimental data, the intrinsic NLD (Eq. 2) was 
multiplied by an energy dependent empirical enhancement factor parameterized as 
\begin{equation}
K_\textrm{\scriptsize{coll}} = 1 + C * exp{[- (U - E_\textrm{\scriptsize{cr}})^2/2\sigma^2]}. \label{eqn3}
\end{equation}
where $C$, $E_\textrm{\scriptsize{cr}}$ and $\sigma$ are the magnitude, peak and width of the enhancement factor, respectively.
At 26 MeV excitation energy, the neutron spectrum has contribution from 1n ($^{168}$Tm) and 2n ($^{167}$Tm) decay channels. But, the higher part of the spectrum ($\geq$ 5 MeV) is totally dominated from the first step decay. Hence, the neutron spectrum was analysed by including the enhancement factor in the NLD of $^{168}$Tm nucleus. 
The extracted parameters for $^{168}$Tm nucleus were $k$ = 8.0 $\pm$ 0.4 MeV, $C$ = 7 $\pm$ 2, $E_\textrm{\scriptsize{cr}}$ = 8.3 $\pm$ 0.5 MeV and $\sigma^2$ = 1.0 $\pm$ 0.3 MeV$^2$. Next, the same parameters were used to explain the $\gamma$ spectra. As can be seen, the $\gamma$  spectrum could be explained below 14 MeV but it was not possible to explain the large yield at 16 MeV by varying the strength of the GDR component (red dotted line in Fig. \ref{fig1}d). 
Therefore, an enhancement was also included in the NLD of $^{169}$Tm nucleus as the high-energy GDR decay will be dominant 
from the first stage of the compound nuclear decay.   
The extracted parameters for $^{169}$Tm nucleus were $C$ = 11 $\pm$ 3, $E_\textrm{\scriptsize{cr}}$ = 9.0 $\pm$ 0.5 MeV and $\sigma^2$ = 1.0 $\pm$ 0.3 MeV$^2$. The extracted GDR centroid energy ($E_\textrm{\tiny{GDR}}$), width ($\Gamma_\textrm{\tiny{GDR}}$) and strength ($S_\textrm{\tiny{GDR}}$) were $E_\textrm{\tiny{GDR1}}$  = 12.1 $\pm$ 0.4 MeV, $\Gamma_\textrm{\tiny{GDR1}}$ = 3.3 $\pm$ 0.6 MeV, $S_\textrm{\tiny{GDR1}}$ = 0.3 $\pm$ 0.04, $E_\textrm{\tiny{GDR2}}$ = 16.0 $\pm$ 0.5 MeV, $\Gamma_\textrm{\tiny{GDR2}}$ = 4.1 $\pm$ 0.7 MeV, $S_\textrm{\tiny{GDR2}}$ = 0.72 $\pm$ 0.05.
We emphasize here that the extracted GDR centroid energies are very similar to the ground state values of $^{165}$Ho (12.2 and 15.8 MeV) measured by livermore group \cite{berm75, atlas} and also to those extracted for $^{166}$Er nuclei (having same deformation) at slightly higher temperature \cite{goss85}. Our result supports the predictions of Brink-Axel hypothesis. 
The estimated deformation from the two GDR peaks is $\beta$ = 0.32 similar to the ground state deformation of Tm nuclei \cite{moll95}.
The bremsstrahlung component, as measured and observed in our earlier experiments at similar beam energy \cite{bala14, supm12},
was parameterized by an exponential function (e$^{-E_{\gamma}/E_0}$) 
where the slope parameter E$_0$ was chosen according to the bremsstrahlung systematics \cite{nif90}.
Interestingly, the enhancement factor used for $^{168}$Tm to describe the neutron spectrum, simultaneously explains the $\gamma$ spectrum between E$_\gamma$ = 7 and 11 MeV. 
This enhancement occurs due to the folding of the low energy tail of the 12.1 MeV GDR component with the 
enhanced level density region after the decay of one neutron populating $^{168}$Tm.
Thus, almost similar enhancement was required in the level density of both $^{168}$Tm and $^{169}$Tm to simultaneously explain the neutron and the GDR spectra.
It is also very interesting to note that no such enhancement in the $\gamma$ spectra was observed in our earlier experiments at similar excitation energies for near spherical nuclei $^{97}$Tc \cite{bala14},  $^{119}$Sb\cite{supm12}, and $^{201}$Tl\cite{dipu12}.    
It needs to be mentioned here that a similar enhancement in NLD was observed in the proton decay from $^{104}$Pd but at much lower effective excitation energy (below 6 MeV) \cite{mitra06}. The enhancement was explained considering pairing re-entrance at high angular momentum \cite{hung15}.
However, the pairing effect does not seem to be the plausible reason for the enhancement in our case, as it has been found to play an important role only below 6 MeV excitation energy and dominant in even-even nuclei \cite{melb99, ozen13, hung15}.
Therefore, the enhancement in NLD for both $^{168}$Tm  and $^{169}$Tm at similar excitation energy primarily appears to be due to the collective enhancement owing to large deformation of Tm nuclei (also observed experimentally via GDR).

\begin{figure}
\begin{center}
\includegraphics[height=7.0 cm, width=8.5 cm]{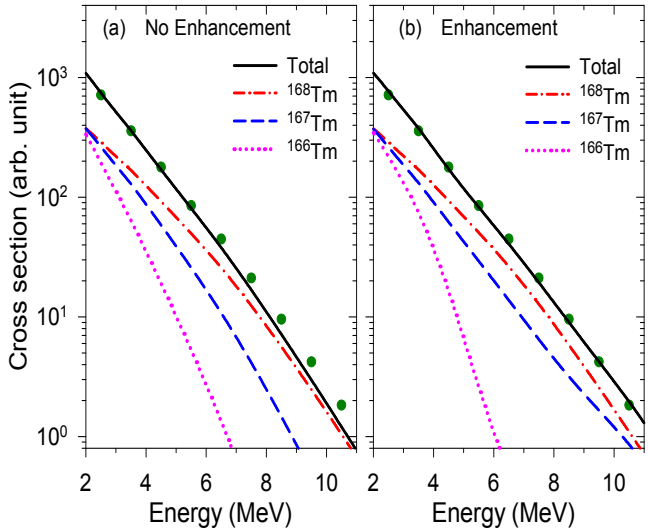}
\caption{\label{fig2} (color online) The symbols represent the neutron spectra (no enhancement in NLD) as calculated from CASCADE with $k$ = 9.5 MeV for the reaction $^4$He(E$_\textrm{\scriptsize{Lab}}$=40 MeV) + $^{165}$Ho studied earlier \cite{prat13}. The same calculation (continuous lines) but with $k$ = 8.0 MeV and (a) no enhancment in NLD of any nuclei and (b) enhancement in NLD of all the three nuclei.}
\end{center}
\end{figure}

\begin{figure}
\begin{center}
\includegraphics[height=12.5 cm, width=8.5 cm]{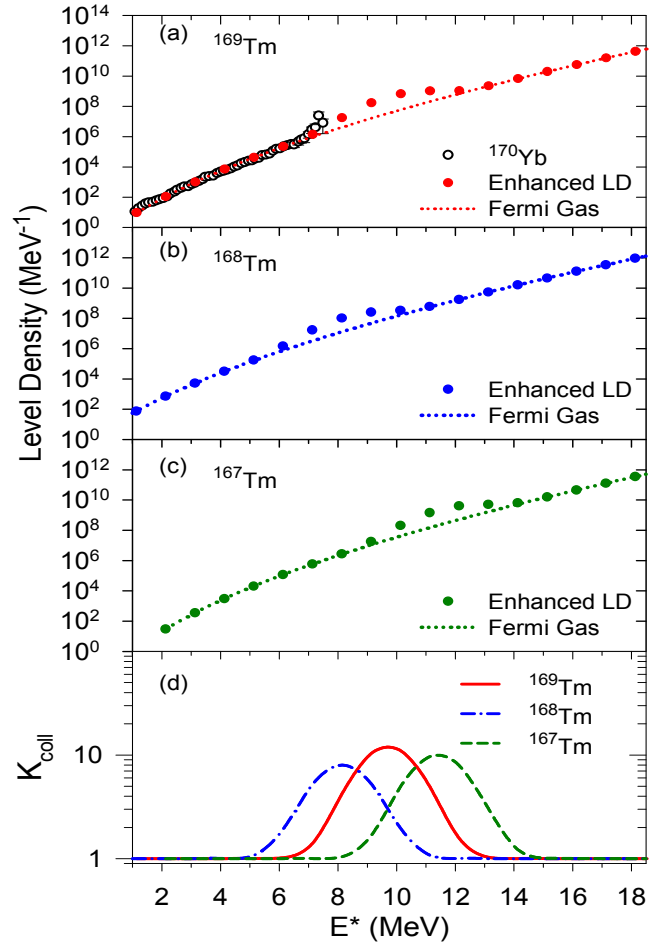}
\caption{\label{fig3} (color online) The enhanced level densities of $^{169}$Tm (a) and $^{168}$Tm (b) at 11$\hbar$ and $^{167}$Tm (c) at 16$\hbar$ are shown as used in the CASCADE. The Fermi gas level density is also displayed for comparison. The level densities are not in absolute scale as they are not normalised to experimental data. The level density of $^{170}$Yb is compared with $^{169}$Tm. (d) The relative enhancement factors extracted as a function of excitation energy  for three nuclei.}
\end{center}
\end{figure}

Recently, a sudden change in the value of $k$ from 8 to 9.5 MeV was obtained for several deformed nuclei indicating the appearance and fadeout of collectivity \cite{prat13, bane17}.  We illustrate how the collective enhancement is manifested through the neutron evaporation spectra when populated in the excitation energy range 32 - 37 MeV. A statistical model calculation with $k$ = 9.5 MeV for the reaction $^4$He (E$_\textrm{\scriptsize{Lab}}$ = 40 MeV) + $^{165}$Ho (performed earlier \cite{prat13}) is shown in Fig. \ref{fig2} (symbols). The same calculation with $k$ = 8.0 MeV (continuous line) is also displayed along with the contributions from different decay steps. As expected, the two calculations are completely different in the higher energy region. However, the two spectrum match very well when collective enhancement is included in the calculation. As can be seen from Fig. \ref{fig2}b, the 1n channel does not see the enhanced region of $^{168}$Tm and is unaffected. Interestingly, the cross section of the 2n channel in the higher energy region increases since it probes the enhanced level density region. Thus, the enhancement factor of $^{167}$Tm  was extracted by fitting the spectra of $k$ = 9.5 MeV (enhancement not included) with $k$ = 8.0 MeV and the enhancement factor. The extracted parameters are $C$ = 9 $\pm$ 3, $E_\textrm{\scriptsize{cr}}$ = 11 $\pm$ 1 MeV and $\sigma^2$ = 1.0 $\pm$ 0.3 MeV$^2$. It was not possible to extract the parameters of 3n channel decay populating $^{166}$Tm since its contribution was very small (Fig. \ref{fig2})   
and same enhancement parameter as $^{167}$Tm was used for the calculation (E$_\textrm{\scriptsize{Lab}}$ = 40 MeV). Thus, the sudden change in the value of $k$ observed in the experiments for deformed nuclei is due to the enhanced cross section of the 2n channel which changes the slope of the neutron spectra. This is compensated in the statistical calculations by changing the $k$ value when enhancement factor is not included. The slope of the neutron spectra is mostly decided by the 1n and 2n decay steps. Therefore, at further higher exciation energy, the first two steps do not see the enhanced level density region and thus, no signature of collective enhancement is observed in the neutron spectra at higher energies \cite{bane17}.

The enhanced level densities, for different Tm nuclei, used in the statistical model calculation are shown in Fig. \ref{fig3}.
An indication of such enhancement in the level density beyond 7 MeV was also seen for $^{170}$Yb \cite{agva04} obtained by the Oslo technique and is compared with $^{169}$Tm in Fig. \ref{fig3}a.       
The collective enhancement factors are also displayed independently in Fig. \ref{fig3}d (on a logarithmic scale) with excitation energy.
Since, neutron evaporation and $\gamma$ ray emission in the statistical model is decided by the ratio of the level density of the daughter nucleus after particle/$\gamma$ ray emission to the compound nucleus, the observed $K_\textrm{\scriptsize{coll}}$ are relative collective enhancement factors.    
The magnitudes of $K_\textrm{\scriptsize{coll}}$ are similar to the  microscopic shell model Monte Carlo calculations for $^{154}$Sm nucleus having similar deformation ($\beta \sim$ 0.27) {\cite{ozen13}. It is also consistent with the prediction in terms of state density of nucleus and its redistribution \cite{grim08}.
The enhancement region for all the three Tm isotopes (having similar ground state deformation \cite{moll95}) is almost same and the collectivity fades away beyond 14 MeV corresponding to the temperature T = 0.82 MeV (Fig. \ref{fig3}d). Interestingly, the deformation is observed directly via the splitting of the GDR strength but the enhanced yield is obtained only for the 16 MeV GDR component. This clearly points that the fadeout of the enhancement is indeed around 14 MeV excitation energy, else an enhancement in 12 MeV GDR component should also have been prominent.
This is also corroborated by the neutron spectrum (Fig. \ref{fig1}c) where the enhancement is observed beyond 6 MeV which corresponds to 12 MeV excitation energy.  
Intriguingly, the result also suggests that the fadeout of the collective enhancement occurs much before the nuclear shape transition from deformed to spherical as predicted by theoretical calculations {\cite{ozen13} and phenomenological estimations {\cite{hans83, jung98}.  One of the reasons for this behavior could be the thermal shape fluctuations ($\Delta\beta$) which increase with the increase in T, as explained earlier for the same fadeout zone for different deformations \cite{bane17}. 
The calculations showed that the nuclear deformation persists at the ground state value up to T $\sim$ 0.8 MeV and then starts the gradual shape change and becomes spherical at T $\sim$ 1.7 MeV. Thus, at around T = 0.8 MeV, the ground state deformation starts to decrease and the thermal fluctuations become large ($\Delta\beta$/$\beta$ = 0.25). This convolutes the static ground state deformation which could lead to the loss of collectivity. Microscopically, the origin of this enhancement in NLD does not come from the levels or states created by deformation.  It appears due to the rearrangement of the levels owing to deformation from higher energy to lower energy which are in the original basis {\cite{grim08}. Hence, when the ground state deformation changes slightly (T $\sim$ 0.8 MeV) and the role of thermal fluctuations becomes large, the energy levels may once again be redistributed leading to the decrease in levels at that particular E* which will appear as loss of collectivity even in the presence of large deformation. 
However, further experimental and theoretical insights are required to understand the details of such a unique behavior.
In deformed nucleus the enhancement also depends on J and K apart from U {\cite{grim13}, and further work is required to see their influence on the enhancement factor.

In summary, we present an experimental evidence of collective enhancement in NLD by measuring the GDR $\gamma$ rays and neutron decays from Tm nuclei. The relative enhancement factors estimated from the simultaneous analysis of  GDR and neutron decays for all the three Tm isotopes are of the order of 10. 
Our technique only measures the apparent change in the enhancement factor and is not sensitive to the magnitude of the vibrational enhancement factor, unless it changes with energy. The experimental result also shows that the collective enhancement fades away beyond 14 MeV excitation energy which is much before the nuclear shape transition from deformed one to spherical predicted by theoretical models.

The authors are thankful to VECC cyclotron operators for smooth running of the accelerator during the experiments.

\end{document}